# Zitterbewegung of Klein-Gordon particles and Landau levels

M.B. Belonenko, N.N. Konobeeva

Volgograd State University

belonenko@volsu.ru

In this paper, we consider the Zitterbewegung (ZB) effect for Klein-Gordon particles induced by a cosmic string, near which curvature effects play a dominant role. An analytical expression for the current is obtained and analyzed.

## 1. Introduction

For the first time, the concept of ZB (trembling motion) was introduced by Schrödinger for free relativistic electrons in vacuum [1] back in 1930. At the same time, interest in this issue with renewed vigor resumed only 15-20 years ago, especially in relation to semiconductors [2, 3].

On the other hand, cosmic strings [4, 5], which are linear topological defects similar to flow tubes in type II superconductors or vortex filaments in superfluid helium, are of great interest [4, 5]. At present, the quantum approach is increasingly being used in the study of cosmic strings, which is associated with progress in the study of various quantum effects [6, 7], including ZB in the time space of a cosmic string [8].

In this paper, we consider bosons. It is known, that charged particles under the action of magnetic fields form Landau levels [9]. Despite recent advances in Landau's bosonic levels [10], it should be noted that they are still poorly understood. For example, the question of the existence of the effect of the trembling motion of bosons at the Landau levels is ignored. But this issue is important for space, astronomical and cosmophysical applications.

The search for ZB for bosons has already been carried out in Ref. [11]. It is shown that trembling movement is not observed in the Foldy-Wouthuysen representation, since position and velocity operators and their transformations into other representations are quantum-mechanical analogs of the corresponding classical variables. Note that the question of the existence of ZB is still debatable. There are even works [12] that refute the observation of a trembling movement in 2010 [13].

In the present paper, the motion of bosons in the vicinity of a cosmic string in a curved space-time is studied for the first time. Thus, we will be able to avoid the above problem with quantum mechanical analogues.

**2. Basic equations**

The relativistic quantum motion of a charged bosonic particle in curved space and in the presence of electromagnetic and scalar potentials is described by the modified Klein-Gordon equation [14]:

$$\left[\frac{1}{\sqrt{-g}}D_\mu\left(\sqrt{-g}g^{\mu\nu}D_\nu\right) - \xi R + \left(M + S(r)\right)^2\right]\phi(x) = 0 \quad (1)$$

where $D_\mu = \partial_\mu + ieA_\mu$, $g = \det(g_{\mu\nu})$, $S(r) = \frac{\eta_C}{r} + \eta_L r$ is the scalar potential, $M$ is the boson mass. Since we are considering the minimum connection, then $\xi = 0$.

The electric current density $j$ can be calculated by the formula [15]:

$$j(x) = \frac{\hbar}{2im}\left[\phi^*(\nabla\phi) - (\nabla\phi^*)\phi\right] \quad (2)$$

The wave function has the form [14]:

$$\phi(x) = \sum exp(i(kz + m\varphi - Et)) \cdot R(r) \quad (3)$$

Substituting (3) into (1) we obtain the radial function:

$$R(r) = \frac{u(r)}{\sqrt{r}} \quad (4)$$

Next, we turn to new dimensionless variables:

$$x = \sqrt{\Delta}r, \Delta = \sqrt{\eta_L^2 + M^2\omega^2} \quad (5)$$

$\omega$ is the particle cyclotron frequency.

Function $u(x)$ can be written in the following form:

$$u(x) = x^{\tilde{\beta}} e^{-0.5x^2} F(x), \quad (6)$$

where $\tilde{\beta} = 0.5 + \frac{|m|}{\alpha}$, the parameter $\alpha<1$ s expressed in terms of the linear mass density of the string $\mu$ as: $\alpha = 1-4\mu$, $F(x)$ is the helper function:

$$F(y) = {_1F_1}\left(\frac{1}{2}\left(\frac{|m|-m}{\alpha}+1\right) - \frac{E^2 - k^2 - M^2}{4M\omega}, \frac{|m|}{\alpha}+1; y\right) \quad (7)$$

where ${_1F_1}(a, b; c)$ is the confluent hypergeometric function.

The energy spectrum $E$ of a particle is given by the formula:

$$E_{k,m,n} = \pm\left\{k^2 + M^2 + 2M\omega\left(2n + 1 + \frac{|m|-m}{\alpha}\right)\right\}^{0.5} \quad (8)$$

*n* is the principal quantum number (*n*=0,1,2…), *m* is the orbital quantum number. The wave packet defined by formula (3) is limited by the sum of *n* from 1 to *N* (N=4, 5, 6), the sum of *m* from 0 to (*n*-1).

## 3. Discussion

The dependence of the current on time according to (2) is shown in Figure 1.

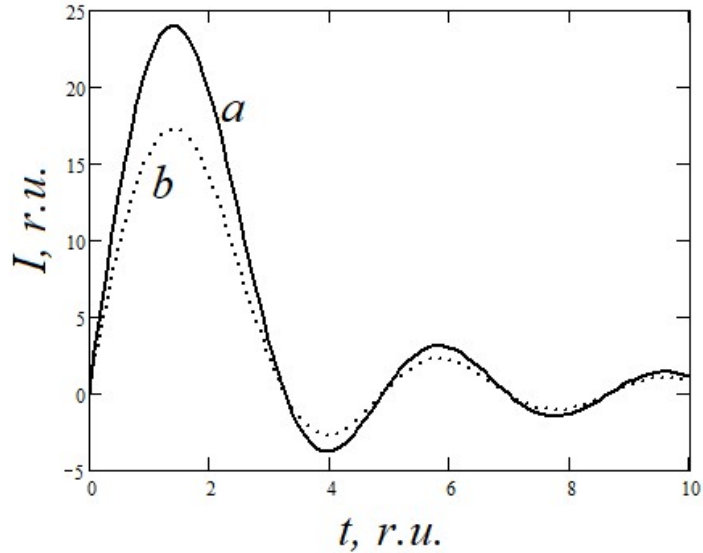

Fig.1. The current versus time: a) *r*=0.1 r.u.; b) *r*=0.2 r.u.

It can be seen from the dependences in Fig. 1 that at large distances from the string, the amplitude of current oscillations caused by the trembling motion of bosons decreases. This behavior is explained by the weakening of the effects associated with the curvature of space-time, with distance from the string.

The dependence of the current on the distance to the event string is shown in Fig. 2 for the selected point in time.

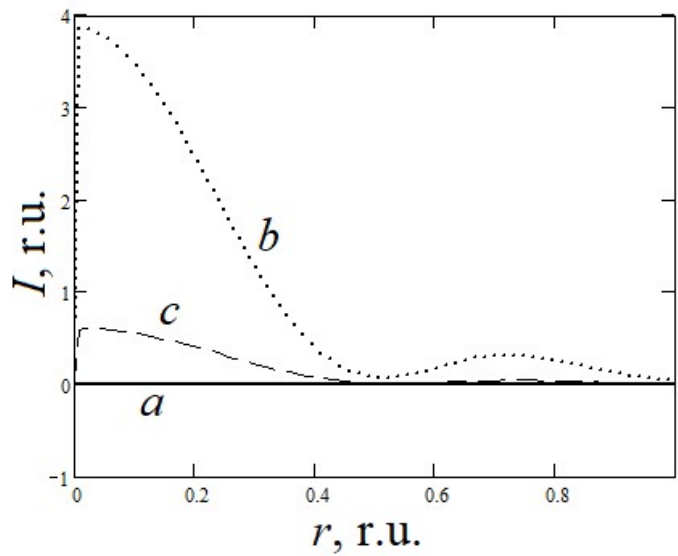

Fig.2. The current versus *r*: a) *t*=0; b) *t*=3; c) *t*=5.

Here, the explanation is related to the behavior of the bosonic wave functions. Namely, their asymptotic behavior near the string. Thus, at small distances, the determining factors are the effects associated with the metric of the cosmic string. At large distances, however, space-time wave packets behave similarly to their behavior in Minkowski space. In addition, the decrease in the modulus of the wave function at large distances according to (4) also plays its role.

The effect of the shape of the wave packet is shown in Fig. 3.

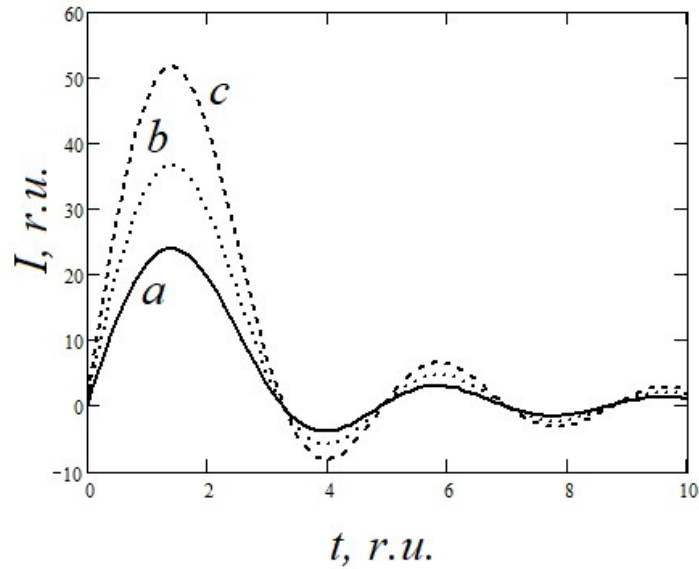

Fig.3. Time dependence of current for different wave packets: a) $N=4$; b) $N=5$; c) $N=6$.

From Fig. 3 it can be seen that the more harmonics in the wave packet, the stronger the Zitterbewegung effect is manifested.

**4. Conclusion**

The following main conclusions can be drawn from the study:

1. The effect of "trembling motion" of bosons in the curved space-time of a cosmic string is discovered.

2. It is shown that the farther we are from the string, the weaker the Zitterbewegung effect becomes, which is associated with a decrease in the wave function with increasing distance.

3. The amplitude of the "trembling motion" can be controlled by the shape of the wave packet.

**5. Acknowledgment**

Authors thank the Ministry of Science and Higher Education of the Russian Federation for the numerical modeling and parallel computations support under the government task (0633-2020-0003).